\documentclass[aps, prl, reprint, twocolumn, superscriptaddress]{revtex4-1}
\usepackage{multirow}
\usepackage{graphicx}
\usepackage{longtable}
\usepackage[utf8]{inputenc}
\usepackage[T1]{fontenc}
\usepackage{epstopdf}
\usepackage{textcomp}
\usepackage[cmyk,dvipsnames]{xcolor} 
\usepackage{amsbsy}
\usepackage{amsmath}
\usepackage{amssymb}
\usepackage{amsfonts}
\usepackage{dsfont}
\usepackage{mathtools}
\usepackage{braket}
\usepackage{array}
\usepackage{gensymb}
\usepackage{placeins}
\usepackage{dashrule}
\usepackage{xcolor}
\usepackage{booktabs}
\usepackage{hyperref}
\usepackage{float}
\usepackage{braket}
\usepackage{ulem}

\definecolor{ao(english)}{rgb}{0.0, 0.5, 0.0}

\makeatletter
  \def\my@tag@font{\normalsize}
  \def\maketag@@@#1{\hbox{\m@th\normalfont\my@tag@font#1}}
  \let\amsmath@eqref\eqref
  \renewcommand\eqref[1]{{\let\my@tag@font\relax\amsmath@eqref{#1}}}
\makeatother

\allowdisplaybreaks

\begin{document}

\title{ 
Turning chiral skyrmion inside out
}

\author{Vladyslav~M.~Kuchkin}
\affiliation{Peter Gr\"unberg Institute and Institute for Advanced Simulation, Forschungszentrum J\"ulich and JARA, 52425 J\"ulich, Germany}
\affiliation{Department of Physics, RWTH Aachen University, 52056 Aachen, Germany}

\author{Nikolai~S.~Kiselev}
 \email{n.kiselev@fz-juelich.de}
 \affiliation{Peter Gr\"unberg Institute and Institute for Advanced Simulation, Forschungszentrum J\"ulich and JARA, 52425 J\"ulich, Germany}

\date{\today}

\begin{abstract}
The stability of two-dimensional chiral skyrmions in a tilted magnetic field is studied.
It is shown that by changing the direction of the field and its magnitude, one can continuously transform chiral skyrmion into a skyrmion with opposite polarity and vorticity.
This turned inside out skyrmion can be considered as an antiparticle for ordinary axisymmetric skyrmion.
For any tilt angle of the magnetic field, there is a range of its absolute values where two types of skyrmions may coexist.
In a tilted field the potentials for inter-skyrmion interactions are characterized by the presence of local minima suggesting attractive interaction between the particles.
The potentials of inter-particle interactions also have so-called fusion channels allowing either annihilation of two particles or the emergence of a new particle.
The presented results are general for a wide class of magnetic crystals with both easy-plane and easy-axis anisotropy.
\end{abstract}

\maketitle

Chiral magnetic skyrmions (Sks) are localized magnetic vortices~\cite{Bogdanov_89}, which can be stabilized in materials with competing the Heisenberg exchange and the Dzyaloshinskii-Moriya interaction (DMI)~\cite{Dzyaloshinskii,Moriya}.
In the most general case, the stability of chiral Sks requires the presence of a potential energy term as the interaction with an external magnetic field and/or the magnetocrystalline anisotropy. The latter plays an important role in the case of thin films and multilayer systems~\cite{Broeder_91}.
Such systems are typically well-described by the two-dimensional (2D) model of chiral magnet, which is also often utilized for so-called quasi-2D crystals -- bulk crystals of particular symmetry allowing DMI between spins contained in specific crystallographic planes, for instance GaV$_4$S$_8$ alloy\cite{Kezsmarki_15}. 
The majority of studies associated with the  stability of Sks~\cite{Ivanov,Bogdanov_1994,Bogdanov_1994JMMM,Bogdanov_1995,Bogdanov_99, Melcher_14,Melcher_17}, their interactions, dynamics and transport properties, have been carried out for the regime when the external magnetic field is applied perpendicularly to the plane of the 2D magnet.
The number of studies related to the case of a tilted field is limited~\cite{Lin_15,W.Wang_15,C.Wang_15,Schmidt_16,Leonov_17_B,S.Zhang_18,Ikka_18,Wan_19,Koide_19}.
These publications are mainly related to Sk lattices, their dynamical properties ~\cite{W.Wang_15,Ikka_18,Koide_19} and phase transitions~\cite{Lin_15,C.Wang_15,Schmidt_16,Leonov_17_B,S.Zhang_18,Wan_19}.
The properties of isolated Sks have been partially discussed in Refs.~\cite{Lin_15,Leonov_17_B}.
In this letter,\! we report\! a number of fundamentally\! new phenomena occurring upon applying\! a tilted\! magnetic field\! to chiral Sks.

We estimate the stability of Sks by means of a direct energy minimization of the micromagnetic functional~\cite{Bogdanov_89}:
\begin{equation}
\mathcal{E} =\! \int\! \Big(  
\mathcal{A}\sum_{i}\left( \mathbf{\nabla} n_i \right)^2 + \mathcal{D}\,w(\mathbf{n}) + U(n_z) \Big)\,t\,\mathrm{d}x\,\mathrm{d}y,
\label{Ham}
\end{equation} 
where  $\mathbf{n}\!=\!\mathbf{M}(\mathbf{r})/M_\mathrm{s}$ is a continuous unit vector field, $M_\mathrm{s}$ is a saturation magnetization, $\mathcal{A}$ and $\mathcal{D}$ are the micromagnetic constants for isotropic exchange and DMI, respectively. 
It is assumed  that magnetization remains homogeneous along the thickness, $t$. 
The DMI term $w(\mathbf{n})$ is defined by combinations of Lifshitz invariants, $\Lambda_{ij}^{(k)}\!=\! n_i\partial_k n_j\!-\!n_j\partial_k n_i$, where $\partial_\beta n_\alpha\!=\!\partial n_\alpha/\partial r_\beta$.
The results presented in this letter are valid for a wide class of chiral magnets of different crystal symmetries with: N\'{e}el-type modulations~\cite{Romming_13,Kez_15,Romming_15} where 
$w(\mathbf{n})\!=\!\Lambda_{xz}^{(x)}\!+\! \Lambda_{yz}^{(y)}$, 
Bloch-type modulations~\cite{Yu_10,Yu_11,Yu_15} where 
$w(\mathbf{n})\! =  \!\Lambda_{zy}^{(x)}\!+\! \Lambda_{xz}^{(y)}\! +\! \Lambda_{yx}^{(z)}\! =\! \mathbf{n}\!\cdot\!\left(\mathbf{\nabla}\!\times\!\mathbf{n}\right)$, and D$_{2\mathrm{d}}$ symmetry~\cite{Nayak_17} where 
$w(\mathbf{n}) \!= \!\Lambda_{zy}^{(x)}\!+\! \Lambda_{zx}^{(y)}$.
The last term in~(\ref{Ham}) represents the potential energy term including uniaxial anisotropy, $U_\mathrm{a}\!=\!K\,(1\!-\! n_z^2)$ and the Zeeman energy -- the interaction with an external magnetic field, $U_\mathrm{Z}\!=\!M_\mathrm{s}\,\mathbf{B}\cdot\mathbf{n}$.
The distances, magnetic fields and energies are given in dimensionless units relative to:
the equilibrium period of helical spin spiral~\cite{helix,Bogdanov_11}, $L_\mathrm{D}\!=\!4\pi\mathcal{A}/|\mathcal{D}|$, the critical field~\cite{Bogdanov_11}, $B_\mathrm{D}\!=\!\mathcal{D}^2/(2M_\mathrm{s}\mathcal{A})$ and the energy of saturated state, $E_0\!=\!2\mathcal{A}t$, respectively. 
The dimensionless magnetic field $\mathbf{h}\!=\!\mathbf{B}/B_\mathrm{D}$ and anisotropy $u\!=\!K/\left(M_{s}B_\mathrm{D}\right)$ are unique control parameters of the system.

\begin{figure*}[ht]
\centering
\includegraphics[width=16cm]{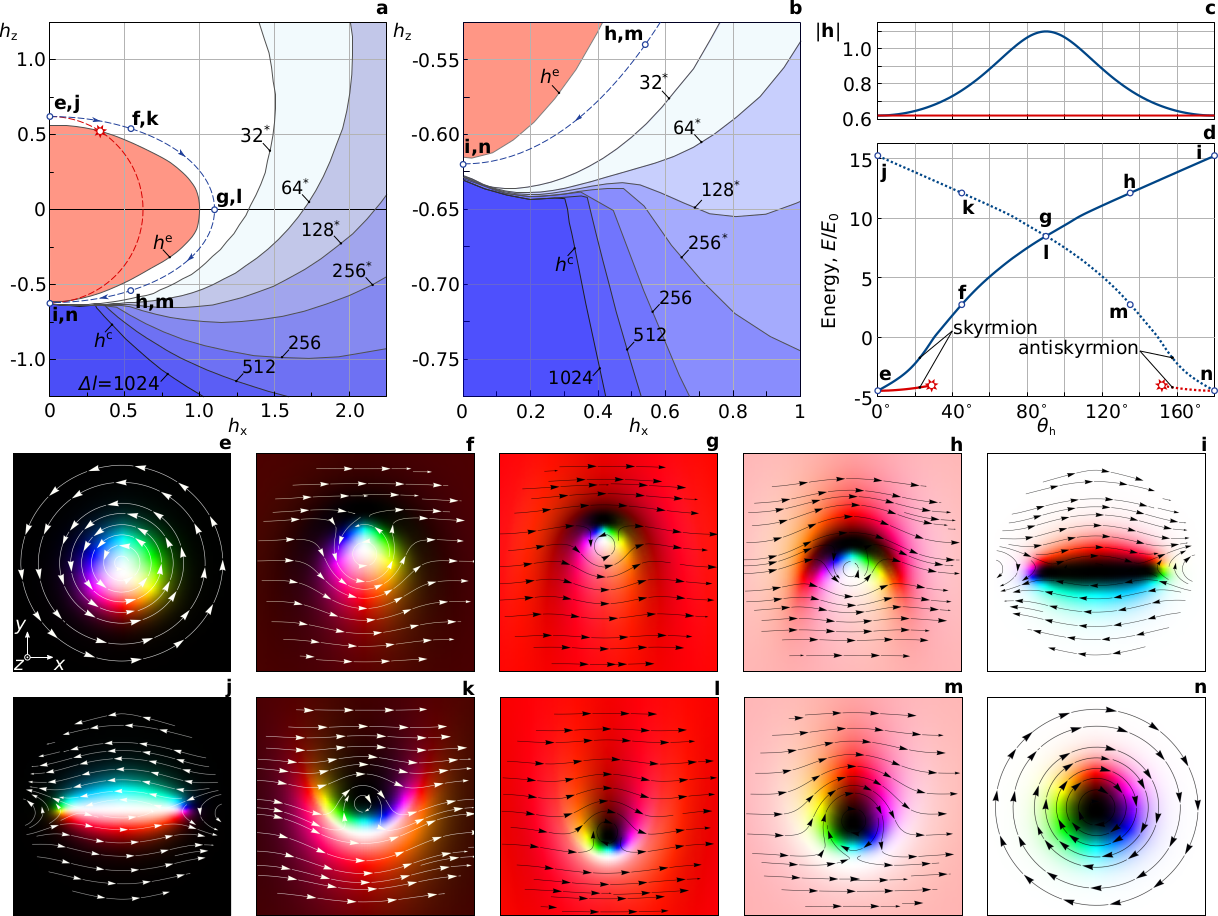}
\caption{~\small \textbf{a,b}
Stability diagram for isolated skyrmion in a tilted magnetic filed, $\mathbf{h}=(h_\mathrm{x},0,h_\mathrm{z})$ at $u\!=\!0$. $h^\mathrm{e}$ and $h^\mathrm{c}$  are the critical fields corresponding to the elliptical instability and the collapse of isolated skyrmion respectively. 
With increasing mesh density $\Delta l$ the region for skyrmion stability extends.
For $\Delta l$ values marked with an asterisk see the main text.
The dashed red line in \textbf{a} corresponds to the case when of $|\mathbf{h}|=0.62$ and does not change with the tilt angle, $\theta_\mathrm{h}$, see red line in \textbf{c}. 
This leads to skyrmion instability at point marked by star symbol in \textbf{a} and \textbf{d}.
The dashed blue line in \textbf{a} and \textbf{b} is the path along which the $|\mathbf{h}|$ changes with $\theta_\mathrm{h}$, see blue line in \textbf{c}. 
At any point $(h_\mathrm{x},h_\mathrm{z})$ along this path there are two stable skyrmion solutions with opposite topological charge $Q$. 
The energies of both solutions as function of $\theta_\mathrm{h}$ are given in \textbf{d}, blue lines: solid line for $Q\!=\!-1$, doted line for $Q=+1$. 
The red lines in \textbf{d} corresponds to the fixed $|\mathbf{h}|\!=\!0.62$.
The spin texture in \textbf{e}-\textbf{i} and \textbf{j}-\textbf{n} represent transient states of two skyrmion solutions with $Q\!=\!-1$ and $Q\!=\!1$ respectively. The images in each row from left to right correspond to $\theta_\mathrm{h}\!=\!0$, $\pi/4$, $\pi/2$, $3\pi/4$, and $\pi$ respectively. 
In \textbf{e}-\textbf{n} the standard color code scheme~\cite{Rybakov_19} is used: black and white correspond to up and down spins respectively, red-green-blue defined by the azimuthal angle with respect to the $x$-axis; the lines and arrows in \textbf{e}-\textbf{n} denote the stream-lines of the in-plane component of the magnetization. For the spin texture see also Supplementary material, Fig. S1~\cite{suppl}.
}
\label{phase_diagram}
\end{figure*}

For direct energy minimization of~(\ref{Ham}) we use a nonlinear conjugate gradient (NCG) method implemented for NVIDIA CUDA architecture and optimized for the best performance by the advanced numerical scheme \textit{atlas}~\cite{Rybakov_15, Rybakov_19}.
We use a fourth-order finite-difference scheme on a regular square grid with periodical boundary conditions~\cite{Rybakov_19}.
The typical size of the simulated domain is $\sim\!10 L_\mathrm{D}\!\times\!10L_\mathrm{D}$ with a mesh density $\Delta l$ given in the number of nodes per $L_\mathrm{D}$, varying from $32$ to $1024$. Most of the results presented below have been obtained for $\Delta l\!=\!52$. In particular cases, in order to approach continuum limit, we employ substantially dense meshes with $\Delta l$ up to 1024 nodes.

We start with the case of purely isotropic system, $u\!=\!0$.
It will be shown below that the discussed phenomena remain valid for a wide range of positive (easy-axis) and \!negative (easy-plane)\! values of the anisotropy. 
The stability diagram for isolated Sk in a tilted magnetic field, $\mathbf{h}\!=\!\!(h \sin \theta_\mathrm{h},0,h\cos \theta_\mathrm{h})$ is shown in Fig.~\ref{phase_diagram}(a).
For any tilt angle, $0\!\leq\!\theta_\mathrm{h}\!\leq\!\pi$ the range of absolute values of the external field, $h$ in which isolated Sk remains stable, is bounded by an elliptic instability field, $h^\mathrm{e}$ from below and by a collapse or blow-up field, $h^\mathrm{c}$ from above. 
The value of the elliptical instability field, $h^\mathrm{e}(\theta_\mathrm{h})$ converges quickly with increase of the mesh density. It does not change significantly for $\Delta l\!\gtrsim\!64$. 
Note, the curve $h^\mathrm{e}$ is not fully symmetric with respect to the horizontal axis $h_\mathrm{z}\!=\!0$.
For instance, in case of perpendicular field, $h_\mathrm{z}^\mathrm{e}\!=\!0.52$ and $-0.62$ for $\mathbf{h}\!\uparrow\uparrow\!\hat{\mathbf{e}}_\mathrm{z}$ and $\mathbf{h}\!\downarrow\uparrow\!\hat{\mathbf{e}}_\mathrm{z}$, respectively.

In contrast to $h^\mathrm{e}$, the collapse field, $h^\mathrm{c}$ does not converge when approaching the continuum limit ($\Delta l\!\rightarrow\!\infty$) as shown in Fig.~\ref{phase_diagram}(a) for different $\Delta l$ values.
The behaviour of the numerical solutions observed here is in line with the results of Ref.~\cite{Melcher_17}, where
it is proven that in the micromagnetic limit, for the particular case of $\theta_\mathrm{h}=0$ and $u\!=\!0$: the stability of skyrmion solution is not bounded from above, even for $h\!\rightarrow\!\infty$.
Following the approach of Ref.~\cite{Melcher_17}, we prove that the above statement can be rigorously extended for the tilt angles at least in the range $0\leq\!\theta_\mathrm{h}\!<\!\pi/3$~\cite{suppl}.

\begin{figure*}[ht]
\centering
\includegraphics[width=15cm]{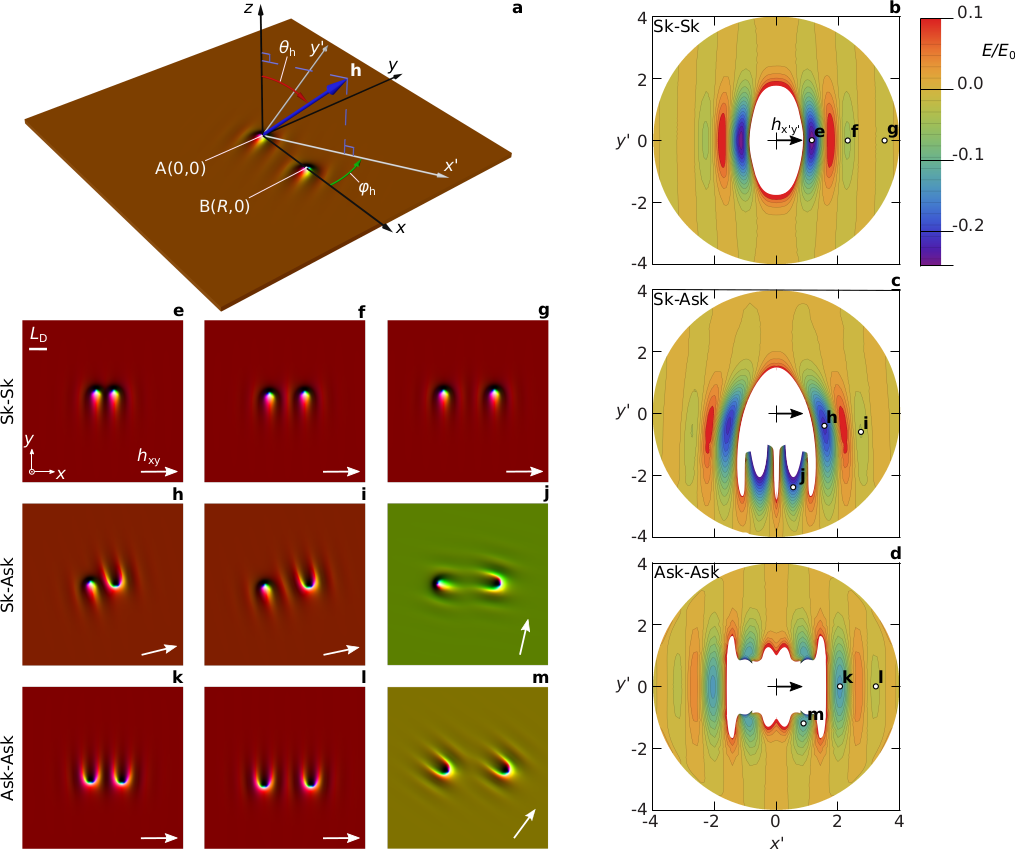}
\caption{~\small \textbf{a} 
Simulated domain of $1024\!\times\!1024$ lattice sites. The points A and B are the positions of two pinned spins with $\mathbf{n}\!=\!(0,0,-1)$ at the center of two interacting skyrmions. \textbf{b}-\textbf{d} are the potential energies of two interacting particles calculated for $\theta_\mathrm{h}\!=\!\pi/3$, $|\mathbf{h}|\!=\!0.8$ ($u\!=\!0$ and $\Delta l\!=\!52$) and given in local coordinate frame $(x^\prime,y^\prime)\!=\!(R\cos\phi_\mathrm{h}, -R\sin\phi_\mathrm{h})$, in which the in-plane component of the field is always pointing along $x^\prime$-axis. The distances $x^\prime$ and $y^\prime$ are given in units of $L_\mathrm{D}$. 
\textbf{b},\textbf{c} and \textbf{d} correspond to skyrmion-skyrmion, skyrmion-antiskyrmion and antiskyrmion-antiskyrmion interactions respectively. 
The spin textures in \textbf{e}-\textbf{g}, \textbf{h}, \textbf{i}, \textbf{k}, \textbf{l} illustrate equilibrium states obtained by energy minimization without spins pinning. These states correspond to the local energy minima marked accordingly in \textbf{b}-\textbf{d}.
The spin textures in \textbf{j} and \textbf{m} were obtained with pinned spins and illustrate the states that are precursory for the fusion of two particles.
All images in \textbf{e}-\textbf{m} have identical size and show only the central part of the simulated domain. The white arrows in the bottom left corner indicate the direction of the in-plane component of the applied magnetic field, $h_\mathrm{xy}$. 
}
\label{potentials}
\end{figure*}

Remarkably, even for fully reversed field, $h_\mathrm{x}\!=\!0$ and $h_\mathrm{z}\!<\!0$, there is a finite range window where Sk solution remains stable, see Fig.~\ref{phase_diagram} (b). 
A continuous transition between the equilibrium Sk solutions shown in (e) and (i) can be achieved by varying the absolute value and the tilt angle of magnetic field, such that $h^\mathrm{e}(\theta_\mathrm{h})\!<\!|\mathbf{h}|\!<\!h^\mathrm{c}(\theta_\mathrm{h})$, see Fig.~\ref{phase_diagram}~(c) and corresponding dashed blue line in (a) and (b).       

Because of continuity of such transition and the fact that states in (e) and (i) have opposite sign of both polarity and vorticity~\cite{polarity&vorticity}, one may conclude that topological charge of these states as well as for all transient states (f),(g), and (h) is identical.
It is also easy to show that for all spin textures in (e)-(h) the topological charge defined by the invariant $Q\!=\!1/(4\pi)\!\int[\mathbf{n}\!\cdot\!(\partial_x \mathbf{n}\!\times\! \partial_y \mathbf{n})]\mathrm{d}x\mathrm{d}y$ equals $-1$.

The symmetry of the problem permits to apply the same analysis to another type of solution, which at $\mathbf{h}\!\downarrow\uparrow\!\hat{\mathbf{e}}_z$ represents an axially symmetric Sk with opposite polarity, see Fig.~\ref{phase_diagram}(n).
The topological charge for the Sk in (n), as well as for all other states in (j)-(m) is $Q\!=\!+1$.
Thereby, the Sk with $Q\!=\!-1$ and the antiskyrmion (ASk) with $Q\!=\!+1$ may coexist at any tilt angle.
Moreover, the energies of Sk and ASk become equal when $\theta_\mathrm{h}\!=\!\pi/2$, see Fig.~\ref{phase_diagram}(d).

The stability of Sk and ASk in a tilted field can be also reproduced in the corresponding spin-lattice model. 
Note that the $h^\mathrm{c}$ curves in Fig.~\ref{phase_diagram}(a,b) for $\Delta l\!=\!32$, 64, 128, and 256 marked with an asterisk have been calculated with the nearest neighbors spin-lattice model where $\Delta l$ has the meaning of a number of lattice sites per $L_\mathrm{D}$~\cite{suppl}.

\begin{figure}[ht]
\centering
\includegraphics[width=8cm]{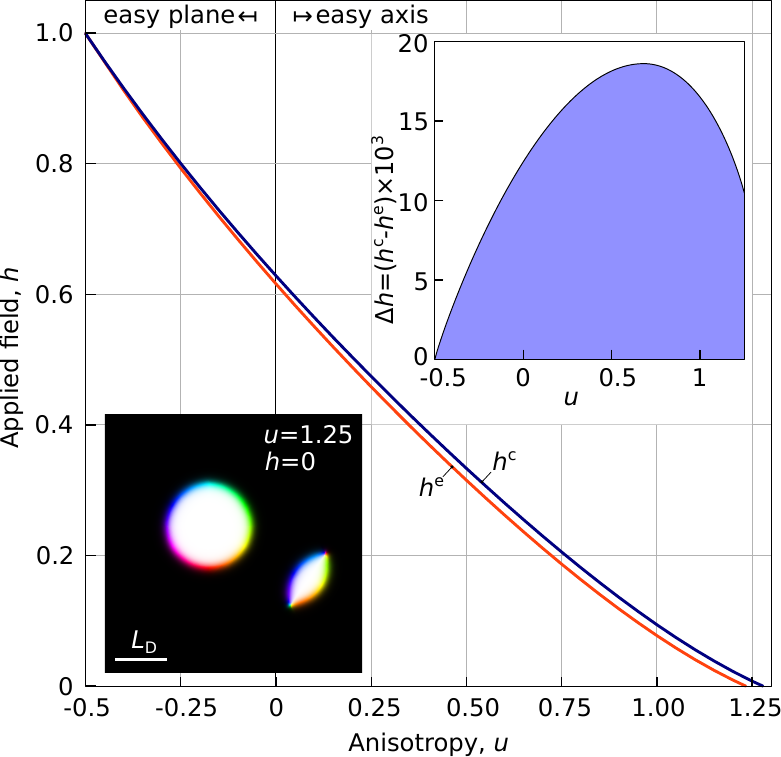}
\caption{\small 
The range of existence for antiskyrmion in perpendicular magnetic field, $\mathbf{h}\!=\!\!(0,\!0,\!h)$ is bounded by the fields of elliptic instability $h^\mathrm{e}$ and collapse, $h^\mathrm{c}$.
The field range $\Delta h$ (right top inset) converges to zero at  Bogomol'nyi point~\cite{Bogomolnyi}, $u\!=\!-0.5$ and $h\!=\!1$. 
The bottom left inset illustrates the coexistence of Sk and ASk without fusion at large distances for the case of strong perpendicular anisotropy and zero field.  
}
\label{h-diagram}
\end{figure}

The interactions between Sks in a tilted field are fundamentally different from those of axisymmetric Sks in a perpendicular field
because of the broken symmetry of their spin texture.
For the calculation of the potentials of the inter-skyrmion interactions presented in Figs.~\ref{potentials}(b-d), we performed the energy minimization on a large size domain with two pinned spins at the centers of the two Sks, as depicted in  Fig.~\ref{potentials}(a).
The white color region in the center of the energy profiles is defined by critical distances, where the interaction energy is either too high or too low. This is attributed to the distortion of the Sks spin texture. 
The calculation of the interaction energies with pinned spins at such small distances has no physical meaning.

The common features of the potentials corresponding to the different pairs of interacting particles in Figs.~\ref{potentials}(b-d) are: i) for any fixed  $\varphi_\mathrm{h}\!\neq\!\pm \pi/2$ the interaction energy between particles oscillates with the distance and changes its sign from negative (attraction) to positive (repulsion); ii) for any pair of particles there are a few local minima corresponding to stable configurations. In fact, we believe that for any $\theta_\mathrm{h}\!>\!0$, there are infinite number of such minima, while their depths as well as the energy barrier between them decay while increasing the distance. The local minima also disappear when $\theta_\mathrm{h}\!\rightarrow\! 0$ or $\pi$~\cite{Lin_15}. 
The ASk interaction presents two remarkable features: an asymmetry of the Sk-ASk potential and the presence of so-called fusion channels.
The latter means that there are specific mutual orientations of the particles leading them to fuse, see Figs.~\ref{potentials}(j) and (m).
Note, this fusion occurs with the conservation of the topological charge, and leads either to the annihilation of the interacting particles or to the emergence of a new particle.
The processes of Sk-ASk and ASk-ASk fusion are presented in Supplementary video files.

The fusion channels are always present on the interaction energy profile involving ASk, even in the perpendicular field.
The latter can be proven via the analysis of the asymptotic behavior of analytical Sk solutions~\cite{suppl}.
The coexistence of Sk and ASk and their ability for fusion both destroy the postulates dominating in earlier works~\cite{Han_10,Nagaosa_review,SOC_review,Fert_review}.
For instance, in Ref.~\cite{Han_10} the authors argued: ``Nor is it possible to fuse a Skyrmion with an anti-Skyrmion to annihilate them since the system under consideration consists only of one species of Skyrmions". 
The statement for the uniqueness of the Sk solution has been disproved in Ref.~\cite{Rybakov_19}. 
Here, we demonstrate the possibility of Sk and ASk fusion. 

The stability of Sks in the whole range of magnetic field tilt angles can be observed not only for isotropic chiral magnet but also for easy-plane and easy-axis anisotropies.
The criterion for this is the stability of the ASk in the perpendicular field. 
The diagram in Fig.~\ref{h-diagram} illustrates a wide range of positive and negative values of $u$ where ASk remains stable. 
Note, the critical point $h\!=\!1,u\!=\!-0.5$ corresponds to so-called Bogomol'nyi point~\cite{Bogomolnyi}, where a large set of skyrmion solutions can be found analytically. This also includes the solution for ASk presented here. 
With increasing $u$, both $h^\mathrm{e}$ and $h^\mathrm{c}$ decrease gradually and for a strong enough uniaxial anisotropy, the stability of Sk and ASk can be achieved even in zero magnetic field, see inset in Fig.~\ref{h-diagram}.

One has to make an important remark regarding previous works reporting the stability of the in-plane Sks similar to those in Fig.~\ref{phase_diagram}(g,l) even at $h=0$, see Refs.~\cite{Ezawa_15,Moon_18,Sitte_19}.
The value of easy-plane anisotropy used in these works corresponds to $u\! \lesssim\!-2$, which is far below the critical value of $u\!=\!-0.5$. 
Indeed, to a certain extent, such solutions can be treated as Sks, although, de facto, they share more similarity to pairs of vortices and antivortices~\cite{suppl}.

In conclusion, in this letter we investigate the stability of 2D chiral magnetic skyrmions in the presence of a tilted magnetic field. 
It is shown that by changing the absolute value of the magnetic field with the tilt angle one can \textit{continuously} transform the axisymmetric skyrmion at $\mathbf{h}\!\uparrow\uparrow\!\hat{\mathbf{e}}_\mathrm{z}$ (or $\mathbf{h}\!\downarrow\uparrow\!\hat{\mathbf{e}}_\mathrm{z}$) into non-axisymmetric skyrmion with opposite polarity and opposite vorticity which remains stable even in fully inverted field $\mathbf{h}\!\downarrow\uparrow\!\hat{\mathbf{e}}_\mathrm{z}$ ($\mathbf{h}\!\uparrow\uparrow\!\hat{\mathbf{e}}_\mathrm{z}$). In other words, one can turn chiral skyrmion inside out. 
It is shown that in the tilted magnetic field the interaction potential of the chiral skyrmions characterized by several local minima revealing attractive interaction  between the particles.
These interactions can be controlled by strength and tilt angle of external magnetic field.
Complex potentials of inter-skyrmion interactions allow not only a formation of various skyrmion clusters but also a fusion of the skyrmions. 
The discussed phenomena are general and applicable for a wide class of crystals with trigonal, tetragonal, and cubic symmetry.

\begin{center}
    {\footnotesize\bf{ACKNOWLEDGMENTS}}
\end{center}
The authors  thank  Filipp Rybakov, Juba Bouaziz and Stefan Bl\"ugel for fruitful discussions during the manuscript preparation.
The authors also thank Filipp Rybakov for providing the software for micromagnetic simulations and Deutsche Forschungsgemeinschaft (DFG) for support through SPP 2137 ``Skyrmionics" Grant No. KI 2078/1-1.

\clearpage
\pagebreak
\onecolumngrid

\setcounter{page}{1}
\setcounter{equation}{0}
\setcounter{figure}{0}
\setcounter{section}{0}
\def\thesection{\arabic{section}}
\renewcommand\thesection{\arabic{section}}
\renewcommand{\thefigure}{S\arabic{figure}}
\renewcommand{\thetable}{S\arabic{section}.\arabic{table}}
\renewcommand{\theequation}{S\arabic{section}.\arabic{equation}}

{\centerline {\bf  \large Supplementary Material for ``Turning chiral skyrmion inside out''}}

\vskip 0.3 true cm

\centerline {Vladyslav~M.~Kuchkin,$^{1,2}$ 
and Nikolai S. Kiselev$^{1}$}

\vskip 0.1 true cm
\centerline {\it $^1$Peter Gr\"unberg Institut and Institute for Advanced Simulation,} 

\centerline {\it Forschungszentrum J\"ulich and JARA, 52425 J\"ulich, Germany}
\centerline {\it $^2$Department of Physics, RWTH Aachen University, 52056 Aachen, Germany}
 
\vskip 0.5 true cm

\setcounter{section}{1}
\setcounter{equation}{0}
\section{S1. Spin lattice model}\label{lattice_model}
The results presented in the main text are based on analysis of micromagnetic model but remain valid also for spin-lattice model: 
\begin{equation}
  E\!=\!- J \!\sum_{\left\langle ij\right\rangle }
   \mathbf{n}_i  \cdot  \mathbf{n}_j - 
  \!\sum_{\left\langle ij\right\rangle } 
  \!\mathbf{D}_{ij} \! \cdot \! [\mathbf{n}_i\! \times\!  \mathbf{n}_j]
  - \! K\!\sum_{i}{n}^2_{\mathrm{z,}i}
  - \! \mu_\mathrm{s} \mathbf{B}_\mathrm{ext}\!\sum_{i}\mathbf{n}_i,   
\label{E_tot_spin}
\end{equation}
where $\mathbf{n}_i = \mathbf{\mu}_i/\mu_\mathrm{s}$ is the unit vector of the magnetic moment at lattice site $i$, $\braket{ij}$ denotes the summation over all nearest-neighbor pairs, $J$ is the Heisenberg exchange constant and $\mathbf{D}_{ij}$ is the Dzyaloshinskii-Moriya vector defined as $\mathbf{D}_{ij}=D\mathbf{r}_{ij}$ with the scalar constant $D$ and the unit vector $\mathbf{r}_{ij}$ pointing from site $i$ to site $j$, $K$ is an uniaxial anisotropy constant, and $\mathbf{B}_\mathrm{ext}$ is an external magnetic field. 

\begin{figure*}[t]
\centering
\includegraphics[width=12cm]{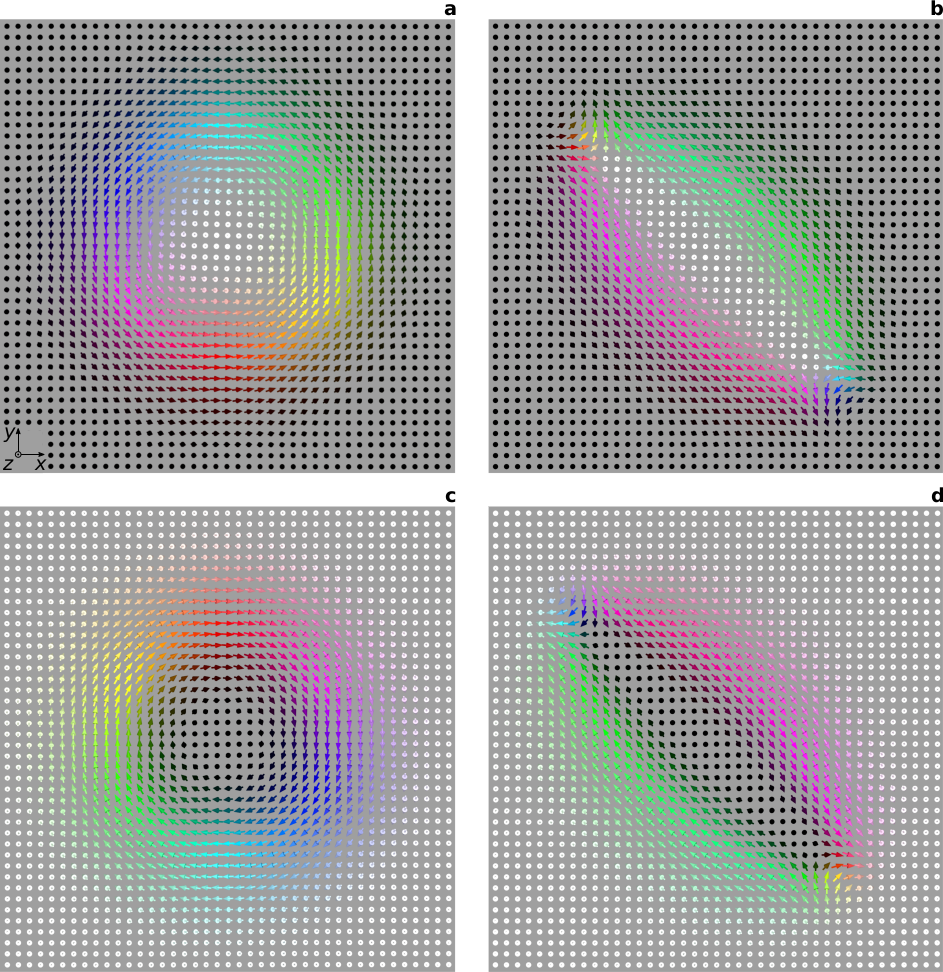}
\caption{~\small Equilibrium spin texture of skyrmions with different polarity, $p$ and vorticity, $\nu$ calculated for the spin lattice model \eqref{E_tot_spin} with the following parameters: $J\!=\!1$, $|D|=0.1963$, $K\!=\!0$, and $\mu_\mathrm{s}B_\mathrm{ext}=0.024$ which correspond to $L_\mathrm{D}\!\approx\!32$ ($L^{*}_\mathrm{D}\!\approx\!32.4$), $u\!=\!0$, $h\!\approx\!0.623$.
The skyrmion in \textbf{a} has $p\!=\!-1$, $\nu\!=\!1$, and topological charge $Q\!=\!-1$, in \textbf{b} $p\!=\!-1$, $\nu\!=\!-1$, $Q\!=\!+1$, in \textbf{c} $p\!=\!+1$, $\nu=1$, $Q\!=\!1$, in \textbf{d} $p=1$, $\nu\!=\!-1$, $Q\!=\!-1$.
The energy minimization has been performed on a square domain of $80\!\times\!80$ lattice sites. 
We use the same color code for spin directions as in the main text.
The black spins are pointed towards the viewer and white spins in the opposite direction.
}
\label{Sk&ASk_in_lattice}
\end{figure*}

The equilibrium period of the spin spiral in model~(\ref{E_tot_spin}) at zero magnetic field has an exact solution~\cite{Yi_2009_supp}:
\begin{equation}
L_\mathrm{D}^{*}=2\pi a/\arctan(D/J).
\label{d_period}
\end{equation}
For weak DMI, $D\!\ll\!J$, it can be approximated with the exact solution of spin spiral period in continuum model:
\begin{equation}
L_\mathrm{D}^{*}\!\approx\! L_\mathrm{D}\!=\!2\pi a J/D.
\label{s_period}
\end{equation}
In Fig.~\ref{Sk&ASk_in_lattice} we show the spin texture of skyrmions corresponding to the case of perpendicular filed. 
It is worth to emphasize that in fully isotropic micromagnetic model the energy of axially nonsymmetric skyrmions (Fig.~1\textbf{i} and \textbf{j}) does not depend on how the semi axes of such elliptic object oriented with respect to the $x$- and $y$-axis. In contrast to this, in the spin-lattice model, there is a significant contribution of anisotropy induced by natural discretization of the system~\cite{Buhrandt_Fritz_supp}. 
A clear manifestation of this effect is that the elliptic shape skyrmion has the lowest energy only when its semi-axes coincide with the diagonals of a square lattice as in Fig.~\ref{Sk&ASk_in_lattice}\textbf{b} and \textbf{d}.

\setcounter{section}{2}
\setcounter{equation}{0}
\section{S2. Energy upper bound for skyrmion in a tilted magnetic field}\label{Energy_bound}

To estimate the stability of a chiral skyrmion in a tilted magnetic field we follow the approach of Ref.~\cite{Melcher_suppl}. In particular, without lost of generality, the functional (1) of the main text can be written in the following reduced form:
\begin{equation}
\mathcal{E}\left(\mathbf{n}\right)=\mathcal{E}_{\mathrm{ex}}\left(\mathbf{n}\right) + \mathcal{E}_{\mathrm{DMI}}\left(\mathbf{n}\right) + \mathcal{E}_{\mathrm{Z}}\left(\mathbf{n}\right)=
\int\!\!\int\! \left[
\frac{1}{2}\left(\nabla\mathbf{n}\right)^{2}\! +\! 2\epsilon
\left(n_\mathrm{x}\frac{\partial n_\mathrm{z}}{\partial y}-n_{\mathrm{y}}\frac{\partial n_\mathrm{z}}{\partial x}\right)\! +\! \frac{\epsilon}{4} \left|\mathbf{n}-\mathbf{e}_\mathrm{z}\right|^{2}
\right]\mathrm{d}x\,\mathrm{d}y,
\end{equation}
where $\epsilon=1/(2h)$. 
We consider an axially symmetric ansatz for skyrmion solution~\cite{Melcher_suppl}.
Next, we model a skyrmion profile in a tilted field by applying a rotation matrix about $y$-axis to all spins:
\begin{equation}
\mathbf{n}=\left(\begin{array}{ccc}
\cos\theta_\mathrm{h} & 0 & \sin\theta_\mathrm{h}\\
0 & 1 & 0\\
-\sin\theta_\mathrm{h} & 0 & \cos\theta_\mathrm{h}
\end{array}\right) \left(\begin{array}{c}
-f'_{R}\left(r\right)\sin\phi\\
f'_{R}\left(r\right)\cos\phi\\
sgn\left(r-1\right)\sqrt{1-\left(f'_{R}\left(r\right)\right)^{2}}
\end{array}\right),
\label{ansatz}
\end{equation}
where the function $f_{R}$ is defined as
\begin{eqnarray}
f_{R}\left(r\right)=
\begin{cases}
\ln\left(1+r^{2}\right),&\mathrm{for }\,\,0\leq r\leq R,\\
c,&\mathrm{for }\,\,r\geq2R,
\end{cases}
\label{ansatz2}
\end{eqnarray}
where $c$ is a positive constant.
The following conditions are applied to the first and second derivatives of the function $f_{R}$:
\begin{equation}
 0\leq f'_{R}\left(r\right)\leq\dfrac{2r}{1+r^{2}},\,\,\,\, 0\leq-f''_{R}\left(r\right)\leq\dfrac{c}{1+r^{2}}, \mathrm{for }\,\,\,r\geq R\gg1. 
\end{equation}
The advantage of such anzatz is that it describes near-core asymptotic behavior of exact skyrmion solution.

In case of the high magnetic field, $\epsilon\ll1$, exchange and Zeeman energy terms are bounded from above:
\begin{equation}
\mathcal{E}_{\mathrm{ex}}\left(\mathbf{n}\right) +  \mathcal{E}_{\mathrm{Z}}\left(\mathbf{n}\right)
\leq4\pi+\pi\epsilon\ln\left(1+R^{2}\right)+c\left(\frac{1}{R^{2}}+\epsilon\right),
\end{equation}

Now let us estimate the bounding energy of DMI.
In the region $r\leq R$ the DMI energy is
\begin{equation}
\mathcal{E}^{*}_{\mathrm{DMI}}\left(\mathbf{n}\right)
=-8\pi\epsilon\frac{R^{2}\left(1+\left(R^{2}-1\right)\cos\theta_\mathrm{h}\right)\cos^{2}(\theta_\mathrm{h}/2)}{\left(1+R^{2}\right)^{2}}.
\label{edmi1}
\end{equation}

In the region $R<r\leq2R$ the DMI energy density is
\begin{equation}
\mathcal{E}^{**}_{\mathrm{DMI}}\left(\mathbf{n}\right)=2\pi\epsilon\!\int_{R}^{2R}\!\!
\frac{\sin^{2}\!\theta_\mathrm{h}\,f'\!\left(1\!-\!f'^{2}\right)\!+\!r\!\left(\sin^{2}\!\theta_\mathrm{h}\!+\!f'^{2}\!\left(\cos\theta_\mathrm{h}\!+\!\cos2\theta_\mathrm{h}\right)\right)f''}{\sqrt{1-f'^{2}}}\mathrm{d}r
\leq\!\int_{R}^{2R}\!\!\frac{2\pi\epsilon \, r\sin^{2}\!\theta_\mathrm{h}}{1+r^{2}}\mathrm{d}r,
\label{edmi2}
\end{equation}
at least when $\cos\theta_\mathrm{h}\!+\!\cos2\theta_\mathrm{h}\geq0$ meaning $\theta_\mathrm{h}\!\leq\!{\pi}/{3}$.
Taking into account (\ref{edmi1}) and (\ref{edmi2}) the upper bound DMI energy, $\mathcal{E}_{\mathrm{DMI}}\!=\!\mathcal{E}^{*}_{\mathrm{DMI}}\!+\!\mathcal{E}^{**}_{\mathrm{DMI}}$ is:
\begin{equation}
\mathcal{E}_{\mathrm{DMI}}\left(\mathbf{n}\right)\leq
-8\pi\epsilon\frac{R^{2}\left(1+\left(R^{2}-1\right)\cos\theta_\mathrm{h}\right)\cos^{2}(\theta_\mathrm{h}/2)}{\left(1+R^{2}\right)^{2}}
+2\pi\epsilon\sin^{2}\theta_\mathrm{h}\ln\frac{1+4R^{2}}{1+R^{2}}.
\end{equation}

Defining $\tilde{\mathbf{n}}=\mathbf{n}\left(\lambda r\right)$, where
$\lambda$ is the rescaling parameter, we get for the total energy of our ansatz solution:
\begin{equation}
\mathcal{E}\left(\tilde{\mathbf{n}}\right)\leq4\pi+\frac{\pi\epsilon}{\lambda^{2}}\ln\left(1+R^{2}\right)+c\left(\frac{1}{R^{2}}+\frac{\epsilon}{\lambda^{2}}\right)
-\frac{8\pi\epsilon}{\lambda}\frac{R^{2}\left(1+\left(R^{2}-1\right)\cos\vartheta_{h}\right)\cos^{2}\frac{\vartheta_{h}}{2}}{\left(1+R^{2}\right)^{2}}
+\frac{2\pi\epsilon\sin^{2}\vartheta_{h}}{\lambda}\ln\frac{1+4R^{2}}{1+R^{2}}.
\end{equation}

Now choosing $R=\dfrac{\left|\ln\epsilon\right|}{\sqrt{\epsilon}}$ and
$\lambda=L\left|\ln\epsilon\right|$ one can write the following inequality for the total energy:
\begin{equation}
\mathcal{E}\left(\tilde{\mathbf{n}}\right)\leq4\pi+
\frac{\epsilon}{\left|\ln\epsilon\right|}\left(-\frac{8\pi}{L}\cos\theta_\mathrm{h}\cos^{2}(\theta_\mathrm{h}/2)+\frac{\pi}{L^{2}}+\frac{2\pi\left(\ln4\right)\sin^{2}\theta_{h}}{L}+\mathcal{O}\left(1\right)\right),
\end{equation}

minimizing this expression with respect to $L>0$ gives
\begin{equation}
L=\left(4\cos\theta_\mathrm{h}\cos^{2}\frac{\theta_\mathrm{h}}{2}-\left(\ln4\right)\sin^{2}\theta_\mathrm{h}\right)^{-1}.
\label{L}
\end{equation}
Note that $L$ in (\ref{L}) remains positive only for angle $\theta_\mathrm{h}\leq{\pi}/{3}$.
This critical angle defines the  limit of our ansatz (\ref{ansatz}). 
Thereby, the energy of skyrmion in external magnetic field tilted by the angle $0\!\leq\!\theta_\mathrm{h}\!\leq\!{\pi}/{3}$ with respect to plane normal is bounded, at least, by the value:
\begin{equation}
\mathcal{E}\left(\tilde{\mathbf{n}}\right)\leq4\pi+
\frac{\epsilon\pi}{\left|\ln\epsilon\right|}\left[-\left(4\cos\theta_\mathrm{h}\cos^{2}\frac{\theta_\mathrm{h}}{2}-\left(\ln4\right)\sin^{2}\theta_\mathrm{h}\right)^{2}+\mathcal{O}\left(1\right)\right].
\end{equation}
It means that the energy of an approximate skyrmion solution is lower than the energy of Belyavin-Polyakov soliton ($4\pi$).
Thereby, the exact skyrmion solution is alos less than $4\pi$ even when  $h\rightarrow\infty$ ($\epsilon\rightarrow0$).

With the simplified ansatz \eqref{ansatz} and \eqref{ansatz2} which does not take into account an asymmetry of the skyrmion in the tilted magnetic field, the above remains true only for $\theta_\mathrm{h}\leq60^\circ$.
Nevertheless, there are no doubts that using a more advanced ansatz one will be able to obtain the higher value of the critical angle which will even better correlate with the results of numerical calculations presented in the main text.

\setcounter{section}{3}
\setcounter{equation}{0}
\section{S3. Analysis of Skyrmion and Antiskyrmion Asymptotics}
To analyse the asymptotic behaviour of  skyrmion and antiskyrmion solutions we solve the variational problem for the micromagnetic functional~(1) in the main text,  
where vector field $\mathbf{n}(x,y)$ is defined by spherical coordinates $\Theta,\Phi$: 
\begin{align}
\mathbf{n}=\left(\begin{array}{c}
\cos\theta_\mathrm{h}\sin\Theta\cos\Phi+\sin\theta_\mathrm{h}\cos\Theta\\
\sin\Theta\sin\Phi\\
-\sin\theta_\mathrm{h}\sin\Theta\cos\Phi+\cos\theta_\mathrm{h}\cos\Theta
\end{array}\right).\label{eq:magnetization}
\end{align}
%
The Euler-Lagrange equations for the functional (1), for radially symmetric solutions is:
\begin{equation}
\begin{cases}
\triangle\Theta+4\pi\sin\Theta\left(\mathbf{n}\cdot\nabla\Phi-\pi h\right)-\dfrac{1}{2}\sin2\Theta\left(\nabla\Phi\right)^{2}=0,\\
\triangle\Phi\sin^{2}\Theta-4\pi\mathbf{n}\cdot\nabla\Theta\sin\Theta+\!\left(\nabla\Theta\cdot\nabla\Phi-2\pi\sin\theta_\mathrm{h}\dfrac{\partial\Theta}{\partial x}\right)\sin2\Theta=0.
\end{cases}\label{LE-equation}
\end{equation}
We assume that boundary conditions obey  $\Theta(0,0)=\pi,\Theta(x,\pm \infty)=\Theta(\pm \infty,y)=0$. 
The asymptotic behavior of the skyrmion solutions of Eq.~\eqref{LE-equation} for $r = \sqrt{x^2+y^2}\rightarrow \infty$ can be written as follows:
\begin{eqnarray}
\Theta&=&{c^{2}}\!/\!{\sqrt{r}}\cdot e^{-2\pi\sqrt{h-\sin^{2}\theta_\mathrm{h}}r},\\ \Phi&=&\pm\left(\phi+{\pi}/{2}\right)+2\pi r\sin\theta_\mathrm{h}\cos\phi+\mathcal{F}\left(\phi\right),
\end{eqnarray}
where plus and minus sign stand for skyrmion and antiskyrmion solutions respectively, $c$ is an arbitrary constant, $\mathcal{F}\left(\phi\right)$ some $2\pi$-periodic function, $r$ and $\phi = \arctan({y}/{x})$ are the radial and angular coordinates in polar coordinate system respectively. 
In the case of perpendicular field, $\theta_\mathrm{h} = 0$ the asymptotic for the skyrmion takes simplified form~\cite{Voronov_asymptot_83, Bogdanov_asymptot_89}:
\begin{eqnarray}
\Theta&=&{c^{2}}\!/\!{\sqrt{r}}\cdot e^{-2\pi\sqrt{h}r},\nonumber\\ \Phi&=&\phi+{\pi}/{2},
\end{eqnarray}
and for antiskyrmion:
\begin{eqnarray}
\Theta&=&{c^{2}}\!/\!{\sqrt{r}}\cdot e^{-2\pi\sqrt{h}r},\nonumber\\ \Phi&=&-\phi-{\pi}/{2}+\mathcal{F}\left(\phi\right),
\end{eqnarray}

\begin{figure*}[t]
\centering
\includegraphics[width=15cm]{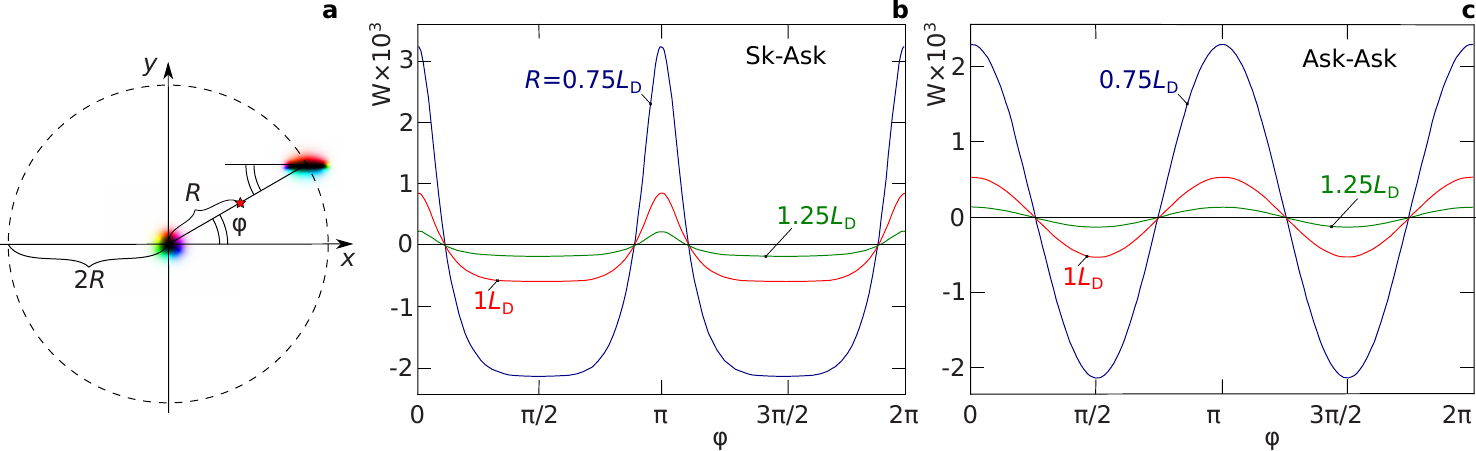}
\caption{~\small \textbf{a} Schematic representation of two interacting particles (skyrmion and antiskyrmion) at the distance $2R$ in the perpendicular magnetic field, $\mathbf{h}=(0,0,0.62)$, $u\!=\!0$.
The mutual orientation of the particles is defined by azimuthal angle $\varphi$. \textbf{b} and \textbf{c} are the potential $W$ calculated at the middle point between particles (star symbol in \textbf{a}) as functions of angle $\varphi$, for skyrmion-antiskyrmion in \textbf{b} and antiskyrmion-antiskyrmion in \textbf{c}.
Note, $W<0$ and $W>0$ correspond to repulsion and attraction respectively.
}
\label{interaction_perp_field}
\end{figure*}

In linear approximation of adding spin waves we get the potential of interaction between skyrmions and antiskyrmions, which equals to the energy density of~(\ref{Ham}) at the middle point between particles, see the star symbol in Fig.~\ref{interaction_perp_field}(a). 
The skyrmion-antiskyrmion and antiskyrmion-antiskyrmion interaction potentials in perpendicular applied magnetic field are shown on Fig.~\ref{interaction_perp_field}~ \textbf{b},\textbf{c}.
In the case of $\theta_\mathrm{h}\neq 0$ interaction energy demonstrates the oscillatory behavior and direct use of the method above is impossible.

\setcounter{section}{4}
\setcounter{equation}{0}
\section{S4. Skyrmion-like states composed of vortices and antivortices. The case of strong easy-plane anisotropy.}

\begin{figure*}[ht]
\centering
\includegraphics[width=16.5cm]{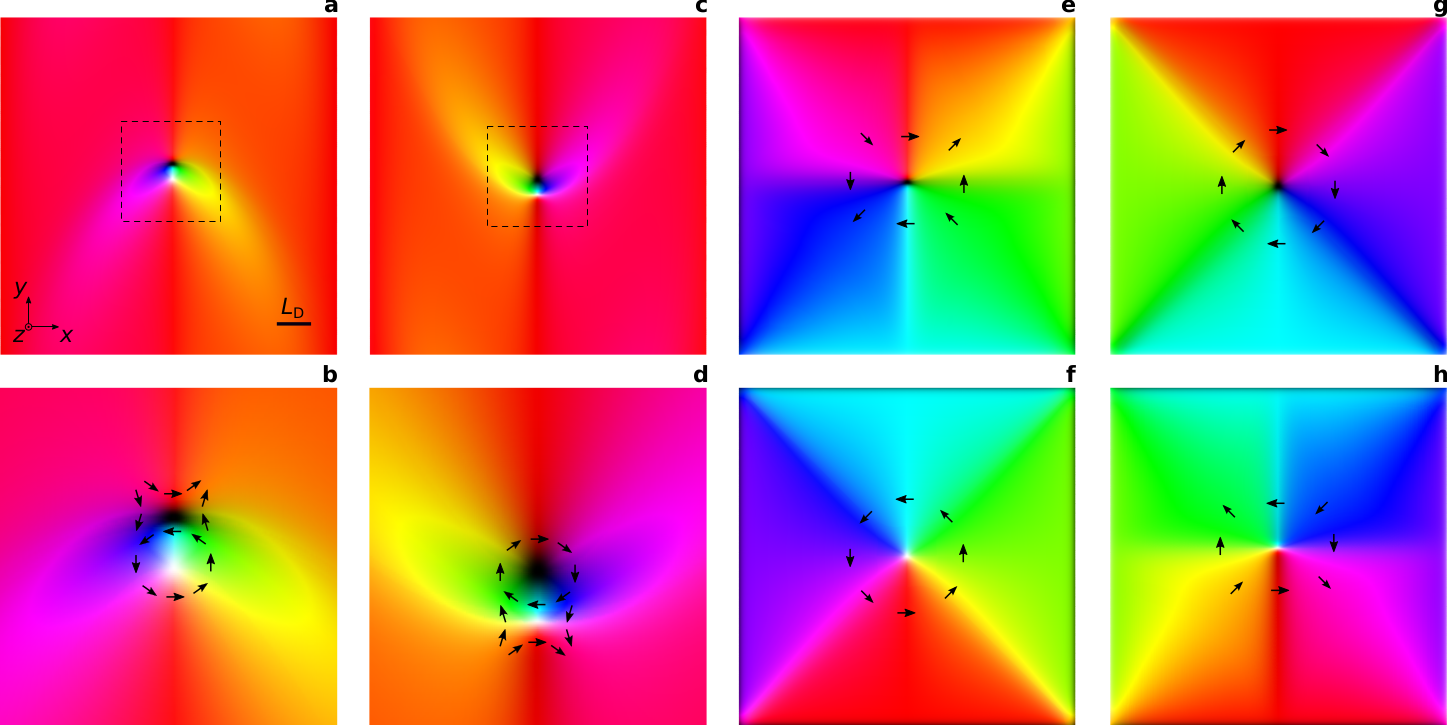}
\caption{~\small 
The equilibrium spin textures obtained by direct energy minimization of functional (1) in the main text for teh case of zero external field, $h\!=\!0$ and strong easy-plane anisotropy, $u\!=\!-2$. We use the parameters corresponding to $\Delta l\!=\!52$ and domain size of  $\approx 10L_\mathrm{D}\times10L_\mathrm{D}$, see the scale bar in \textbf{a}.
The pairs of vortex and antivortex with total topological charge, $Q\!=\!-1$ in \textbf{a} and $Q\!=\!+1$ in \textbf{b}. 
A zoomed area indicated by dashed square in \textbf{a} and \textbf{c} are shown in \textbf{b} and \textbf{d} respectively.
The pairs of vorticies and antivortices in \textbf{a} and \textbf{c} are calculated with periodical boundary conditions in $xy$-plane. 
\textbf{e}-\textbf{h} show isolated vortices and antivortices with different polarity, $p$ and vorticity $\nu$:
$p\!=\!-1$, $\nu\!=\!-1$ (\textbf{e}),
$p\!=\!1$, $\nu\!=\!1$ (\textbf{f}),
$p\!=\!-1$, $\nu\!=\!1$ (\textbf{g}), and 
$p\!=\!1$, $\nu\!=\!-1$ (\textbf{h}).
Blak arrows indicate the in-plane magnetization.
}
\label{V&AV1}
\end{figure*}

In the case of strong easy-plane anisotropy, $u\ll0$ the energy minimization of the states composed of skyrmion or antiskyrmion leads to an appearance of the textures, which are morphologically very similar to that discussed in the main text of the manuscript, compare the states in  Figs.~\ref{V&AV1}\textbf{a} and \textbf{c} to the states depicted in the Figs.~1\textbf{g} and \textbf{l}. 
Such states can be stabilized in the system even without applying an external field~\cite{Ezawa_15_supp,Moon_18_supp,Sitte_19_supp}. Similar to skyrmions these states may have positive or negative topological charge and exhibit particle like properties meaning that they can move and interact with each other~\cite{Ezawa_15_supp,Moon_18_supp,Sitte_19_supp}.
In particular, at small distances the states shown in Figs.~\ref{V&AV1}\textbf{a} and \textbf{c} attract each other which in turn leads to their annihilation. 
Moreover, the particles with identical topological charge attract each other and may form clusters mainly in form of chains of particles.
These two facts, indeed, make a lot of similarity between the objects stabilized at strong easy-plane anisotropy and skyrmions in a tilted field.
On the other hand, there are strong arguments supporting the statement that these objects possess more similarities to pairs of vortex and anivortex rather than skyrmions.

Below we show the examples of different textures appearing in case of strong easy-plane anisotropy. 
These representative textures were obtained by direct energy minimization for the functional (1) with  $|\textbf{h}|=0$, $u\!=\!-2$ and discretization $\delta l\!=\!52$. For comparison, the micromagnetic parameters used in previous works reporting the study of so-called \textit{in-plane} skyrmions corresponds to the following anisotropy: $u\!=\!-1.96$~Ref.~\cite{Ezawa_15_supp}, $u\!=\!-2.00$~Ref.~\cite{Moon_18_supp}, and 
in~Ref.~\cite{Sitte_19_supp} for different set of parameters $u\!=\!-2.88$, $-3.33$, and $-4.67$.

\begin{figure*}[ht]
\centering
\includegraphics[width=16.5cm]{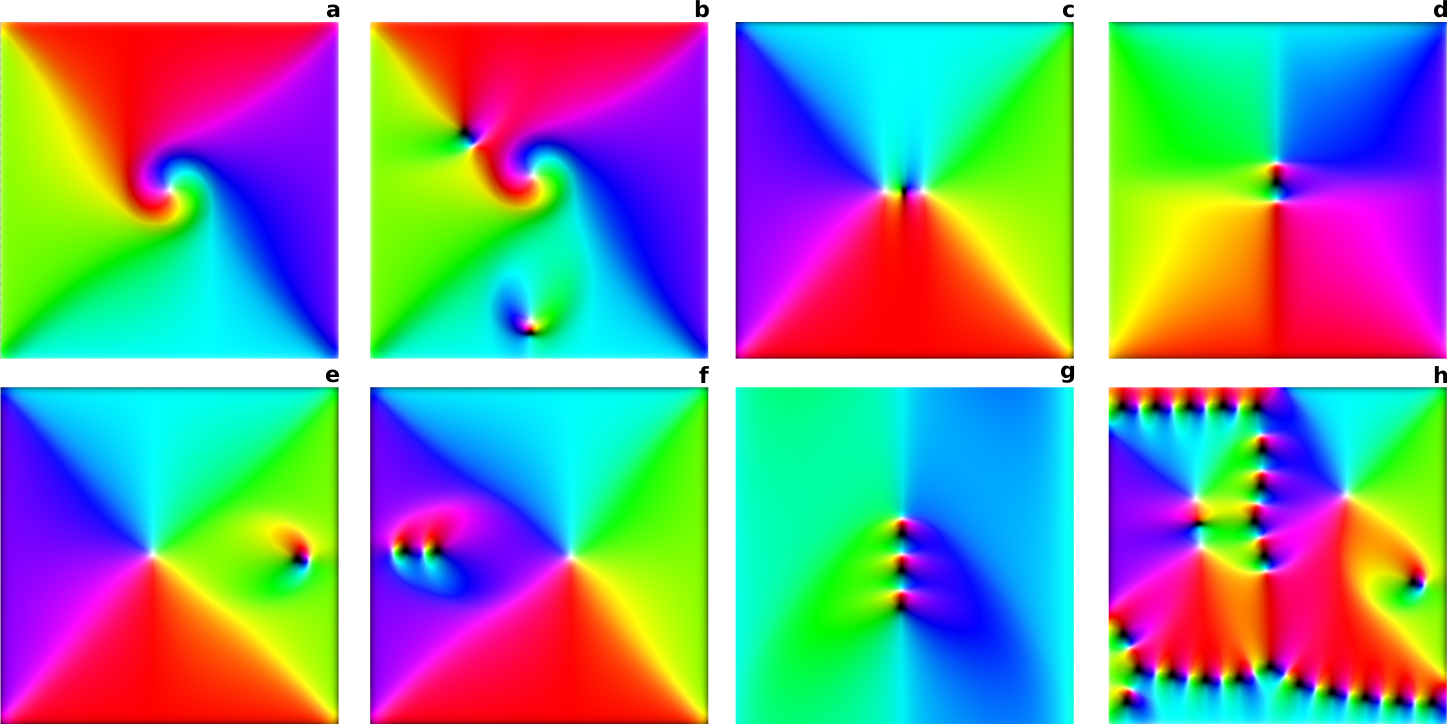}
\caption{~\small Various stable spin configurations in the square shape domain with open boundsry conditions obtained by energy minimization with the same parameters as in Fig.~\ref{V&AV1}. 
\textbf{a} A vortex with $p\!=\!1$, $\nu\!=\!1$, the same as in Fig.~\ref{V&AV1}\textbf{f} but with other type of modulations at the edges. 
\textbf{b} The same vortex as in \textbf{a} coexisting with tow different vortex-antivortex pairs.
\textbf{c} A triplet state composed of two vortices and one antivortex.
\textbf{d} A triplet state composed of two antivortices and one vortex.
\textbf{e} An example of a state composed of three different types of vortices, where mutually attractive vortex and antivortex form a localized pair on the right  with topological charge of the pair $Q\!=\!+1$. The vortex on the left is repelled by that pair.
\textbf{f} Another example of the state composed of three different types of vortices. Two vortices and two antivorticies form a chain on the left with topological charge $Q\!=\!-2$.
\textbf{g} A chain of alternating vortices and antivortices.
\textbf{h} An example of a complex state obtained after full energy minimization starting from random spin distribution. The state contains all four types of vortices and antivortices, see Figs.~\ref{V&AV1}\textbf{e}-\textbf{h}.
The state composed of one isolated vortex in the top right quadrant, pairs, triplets, and chains of vortices and antivortices. 
}
\label{V&AV2}
\end{figure*}

First of all it is easy to show that the states in Figs.~\ref{V&AV1}\textbf{a} and \textbf{c} can be decomposed into stable alone states. 
Indeed, Figs.~\ref{V&AV1}\textbf{e}-\textbf{h} illustrate two types of vortices and antivortices with different polarity of the cores. 
It is well known that isolated vortex can not be stabilized in an infinite sample (or in the domain with periodical boundary conditions) and only a pair of vortex and antivortex can do that. Because of that, in order to visualize such stable isolated vortices in Figs.~\ref{V&AV1}\textbf{e}-\textbf{h}, we use finite size domain with open boundary conditions. 
It is easy to see that the state in \textbf{b} (\textbf{d}) can be decomposed into pair of antivortex in \textbf{e} (\textbf{h}) and vortex in \textbf{f} (\textbf{g}).
Note that due to the presence of DMI the vortex state in \textbf{f} and \textbf{g} have energy lower than that of antivortices in \textbf{e} and \textbf{h}.
Isolated vorticies are also more stable than antivorticies which is illustrated by Fig.~\ref{V&AV2}\textbf{a} where vortex state remains stable even in the case of not favorable edge modulations. In contrast to this the antivortices are unstable at these conditions. 
For the range of uniaxial anisotropy value $u>-0.5$ discussed in the main text of the manuscript the vortices shown in \textbf{e}-\textbf{h} are unstable in the whole field range.
In other words, at these conditions, the skyrmion is a single object which cannot be decomposed into more elementary particles. 

\vspace{0.5cm}
Beside the stability of isolated vortices in case of strong easy-plane anisotropy the vortices and antivortices may also form more complex textures, see Fig.~\ref{V&AV2}.
In particular, vortices can appear not only in form of isolated vortices and pairs as in Fig.~\ref{V&AV2}\textbf{a} and \textbf{b} but also can be stable as a triplet state, quadruplet state, \textit{etc}.
Figures.~\ref{V&AV2}\textbf{c} and \textbf{d} illustrate some stable triplet states.
Note, the vortex-antivortex triplets may appear as a localized bound state only when composed of two types mutually attracting objects.
Only the objects with opposite polarity and opposite vorticity exhibit attractive interactions.
Thereby, only vortex and antivortex can attract each other and only when their polarities are opposite, see Figs.~\ref{V&AV1}\textbf{f} and \textbf{e} for such vortex and antivortex respectively and Figs.~\ref{V&AV1}\textbf{g} and \textbf{h} for another pair of the vortex in and antivortex. 
The interaction of vortex and antivortex with identical polarity always lead to their annihilation.
The pair of two vortices and the pair of antivortices with opposite polarities always characterized by repulsion.
Figure~\ref{V&AV2}\textbf{e} illustrates the state composed of three types of objects with a different character of the interactions: a vortex with positive polarity, vortex with negative polarity, and an antivortex with positive polarity.
The coupled pair of vortex-antivortex on the right side of the figure and the vortex on the left side repel each other.
A similar configuration with a larger number of vortices and antivortices is shown in Fig.~\ref{V&AV2}\textbf{f}.

\vspace{0.5cm}
A typical spin texture after full energy minimization from random distribution is shown in Fig.~\ref{V&AV2}\textbf{h}.
Remarkably, all four types of vortices depicted in Fig.~\ref{V&AV1}\textbf{e}-\textbf{h} are present in Fig.~\ref{V&AV2}\textbf{h}.
One may conclude that such configurations can be quantified only by the number of vortices and antivortices.
Thereby, in the case of strong easy-plane anisotropy,  these four types of vortices and antivortices are the countable objects that describe the state of the system.

\end{document}